\documentclass[reprint,superscriptaddress,pra]{revtex4-1}
\pdfoutput=1
\usepackage{amsmath,amssymb,graphicx,xcolor,units,amsfonts}
\usepackage{color}
\usepackage{verbatim}
\usepackage{tikz}
\usepackage{siunitx}
\usepackage{titlesec}
\usepackage{mathtools}
\usepackage{physics}
\usepackage{bm}
\usepackage{setspace}
\usepackage{hyperref}
\hypersetup{colorlinks,allcolors=blue}
\usepackage[]{siunitx}

\newcommand{\beginsupplement}{%
        \setcounter{table}{0}
        \renewcommand{\thetable}{S\arabic{table}}%
        \setcounter{figure}{0}
        \renewcommand{\thefigure}{S\arabic{figure}}%
     }

\begin{document}

\title{Magnetic solitons in an immiscible two-component
Bose-Einstein condensate}

\author{Xiao Chai}
\email{xchai@gatech.edu}
\affiliation{School of Physics, Georgia Institute of Technology, 837 State St, Atlanta, Georgia 30332, USA}

\author{Li You}
\affiliation{State Key Laboratory of Low Dimensional
Quantum Physics, Department of Physics, Tsinghua
University, Beijing 100084, China}
\affiliation{Frontier Science Center for Quantum Information, Beijing, 10084, China}

\author{Chandra Raman}
\affiliation{School of Physics, Georgia Institute of Technology, 837 State St, Atlanta, Georgia 30332, USA}

\begin{abstract}
We investigate magnetic solitons in an immiscible binary
Bose-Einstein condensate (BEC), where the intraspecies
interactions are slightly weaker than the interspecies
interactions. While their density and phase profiles are
analogous to dark-bright solitons, other characteristic
properties such as velocities, widths, total density
depletions, and in-trap oscillations are different. In
the low velocity regime, a magnetic soliton reduces to a
traveling pair of magnetic domain walls. Collisional
behaviors of the solitons are also briefly discussed. We
further demonstrate that these solitonic states can be
realized in a quasi-one-dimensional (quasi-1D) spin-1
ferromagnetic BEC with weak spin interaction, e.g., a
$\prescript{87}{}{\mathrm{Rb}}$ BEC.
\end{abstract}
\maketitle

\section{Introduction}
Solitons are stable and localized topological excitations
in nonlinear systems. Their stability comes from the
combined actions of dispersion and nonlinearity. Solitons
exist in various physical systems, such as shallow water
\cite{Camassa.Holm.1993}, optical fibers
\cite{kivshar2003optical}, gravitational systems
\cite{Carr.Verdaguer.1983}, solid state materials
\cite{Parkin.Thomas.2008}, and ultracold atomic quantum
gases
\cite{Burger.Lewenstein.1999,Khaykovich.Salomon.2002}.
Systems of ultracold quantum gases stand out as they
provide controllable platforms for solitons, and the rich
internal structures of ultracold atoms facilitate
multi-component solitons, namely vector solitons.
Previous studies of vector solitons in ultracold gases
are mostly confined to the Manakov regime
\cite{Manakov.Manakov.1974}, with equal intra- and
interspecies interaction strengths. Numerous soliton
solutions have been obtained, including dark-bright
solitons \cite{Busch.Anglin.2001,Becker.Sengstock.2008}
in two-component BECs and dark-bright-bright solitons
\cite{Nistazakis.Carretero-Gonzalez.2008,Kevrekidis.Frantzeskakis.2016,Bersano.Kevrekidis.2018,Lannig.Oberthaler.2020}
in three-component BECs.

The Manakov limit, however, constitutes an approximation
for ultracold atomic gases, which is valid provided the
spin-dependent or magnetic dynamics are sub-dominant. In
more realistic two-component Bose systems, the
intraspecies interaction $g_{11},~g_{22}$ and
interspecies interactions $g_{12}$ are usually unequal,
so that quantum magnetism can play a role. In the
immiscible regime, $\delta g \equiv g -g_{12} < 0$ with
$g = \sqrt{g_{11}g_{22}}$, phase separation
\cite{Timmermans.Timmermans.1998,Hall.Cornell.1998}
happens spontaneously and magnetic domain walls
\cite{Coen.Haelterman.2001,Yu.Blakie.2020}, a type of
static vector soliton, emerge as a result of modulation
instability
\cite{Miesner.Ketterle.1998,Kasamatsu.Tsubota.2004}. In
the miscible regime where $\delta g >0$, magnetic
solitons, a special type of traveling soliton decoupled
from the density dynamics, have been proposed recently
\cite{Qu.Stringari.2016}. Magnetic solitons are
dispersion free spin density excitations propagating on
top of a balanced spin background. Recent experiments
indicate that magnetic solitons can be embedded in spin-1
antiferromagnetic BECs of sodium atoms
\cite{Chai.Raman.2020,Farolfi.Ferrari.2020,Chai.Raman.2020jrk}.
Numerical studies \cite{Fujimoto.Ueda.2019} further
reveal the existence of correlations between the
non-equilibrium spinor dynamics and magnetic solitons.

In this study, we report on the discovery of another type
of traveling soliton in the immiscible regime, which can
be considered as the counterpart of the magnetic solitons
in the miscible regime \cite{Qu.Stringari.2016}. Their
properties and existence depend crucially on $\delta g$.
For consistency with earlier conventions, we will also
refer to the traveling solitons we study here as
\textit{magnetic solitons}. Similar to
Ref.~\cite{Qu.Stringari.2016}, in this work we restrict
to the limit $\abs{\delta g} \ll g$ such that the spin
and density dynamics are decoupled and the total density
can be safely assumed as a constant (see
Ref.~\cite{Congy.Pavloff.2016} and the Supplementary
Material \cite{sup} for more detailed discussion). In
reality, this condition is easily fulfilled in a
$\prescript{87}{}{\mathrm{Rb}}$ BEC where $\abs{\delta
g}/g \approx 0.0093$ \cite{Kempen.Verhaar.2002} for a
system composed of two hyperfine states $\ket{F=1,m=\pm
1}$. To our knowledge, solitons in the immiscible regime
have only been explored as static solutions
\cite{Coen.Haelterman.2001,Yu.Blakie.2020}, or as
variants of the dark-bright solitons
\cite{Busch.Anglin.2001,Alotaibi.Carr.2017,Katsimiga.Schmelcher.2017}.

\section{Formalism and Solution}
For a 1D binary BEC, its mean-field equations of
motion can be obtained from the Lagrangian density,
\begin{equation}
    \mathcal{L} = \sum_{j=1}^2 \frac{i\hbar}{2}
    \left(\psi_j^*\pdv{\psi_j}{t}-\psi_j\pdv{\psi_j^*}{t}\right)
    -\mathcal{E},
    \label{eq:lag1}
\end{equation}
where $\psi_j(z,t)$ is the $j$-th component condensate
wave function with $j=1,2$, and $z,t$ are space and time
coordinates, respectively. $\mathcal{E}$ is the energy
density given by
\begin{equation}
    \mathcal{E} = \sum_{j=1}^2 \left(\frac{\hbar^2}{2M}
    \abs{\pdv{\psi_j}{z}}^2 + \mathcal{V}\abs{\psi_j}^2 +
    \sum_{l=1}^2 \frac{g_{jl}}{2}
    \abs{\psi_j}^2\abs{\psi_l}^2\right),
\end{equation}
with $M$ the same atomic mass of both components, and
$\mathcal{V}(z)$ the trapping potential. We will
focus on the parameter regime where
$g_{11}=g_{22}=g=g_{12}+\delta g=g_{21}+\delta g$ with
$\delta g<0$. The wave functions can be parametrized as
\begin{equation}
    \mqty(\psi_1\\\psi_2) = \sqrt{\mathfrak{n}}
    \mqty(\cos{(\theta/2)}e^{i\phi_1}\\
    \sin{(\theta/2)}e^{i\phi_2}),
\end{equation}
where $\theta(z,t),\phi_j(z,t)$ are real and $\mathfrak{n}(z,t)>0$. In the following discussion we
\textit{assume} the total density
$\mathfrak{n}(z,t)$ is a constant $n$. To search for
traveling soliton solutions with a constant velocity $V$,
we write $\theta(z,t) = \theta(z-Vt)$ and $\phi_j(z,t) =
    \phi_j(z-Vt)$. Then in the uniform case with
$\mathcal{V}=0$, the Lagrangian (\ref{eq:lag1}) can be
expressed as
\begin{align}
    \frac{\mathcal{L}}{n M V_s^2} = & \frac{1}{16}\cos{2\theta} - \frac{1}{8}
    (\partial_{\zeta}\theta)^2 \nonumber                                      \\
                                    & +
    \frac{1}{2}U(1+\cos{\theta})\partial_{\zeta}\phi_1 -
    \frac{1}{4}(1+\cos{\theta})(\partial_{\zeta}\phi_1)^2
    \nonumber                                                                 \\
                                    & +
    \frac{1}{2}U(1-\cos{\theta})\partial_{\zeta}\phi_2 -
    \frac{1}{4}(1-\cos{\theta})(\partial_{\zeta}\phi_2)^2,
\end{align}
where $\zeta = (z-Vt)/\xi_s$, $U=V/V_s$ are the
normalized moving coordinate and velocity, respectively.
$\xi_s=\hbar/\sqrt{2Mn\abs{\delta g}}$ is the spin
healing length and $V_s=\sqrt{2n\abs{\delta g}/M}$ is in
fact the maximum speed of the soliton, as it will become
clear later. Our definition for $\xi_s$ differs from the
choice of Ref.~\cite{Kawaguchi.Ueda.2012}. We also omit
constant terms in $\mathcal{L}$ which do not
contribute to the dynamics.

Due to the immiscible nature the background of the
soliton is fully spin-polarized, which means only one
spin component (e.g., the component 1) exists at infinity
and the other spin component is localized. Thus we impose
the following boundary conditions for $\theta$,
\begin{equation}
    \theta=\partial_{\zeta}\theta
    = 0,~\mathrm{at}~\zeta \rightarrow \pm \infty.
    \label{eq:bc1}
\end{equation}
When a global flux is absent for the component 1, the
boundary condition for $\partial_\zeta \phi_1$ is given
by
\begin{equation}
    \partial_{\zeta} \phi_1 = 0,~\mathrm{at}~
    \zeta \rightarrow \pm \infty.
    \label{eq:bc2}
\end{equation}
No restriction for $\partial_{\zeta}\phi_2$ is
supplied at infinity because the component 2 has no
population at infinity.

The variation of the Lagrangian with respect to $\phi_1$
gives
\begin{equation}
    \partial_{\zeta}\left\{-U\cos{\theta}+(1+\cos{\theta})
    \partial_{\zeta}\phi_1 \right\}=0.
\end{equation}
Applying the boundary conditions (\ref{eq:bc1}) and (\ref{eq:bc2}) we find
\begin{equation}
    \partial_{\zeta}\phi_1 = - U
    \frac{1-\cos{\theta}}{1+\cos{\theta}}.
    \label{eq:phi1}
\end{equation}
The variation with respect to $\phi_2$ gives additionally
\begin{equation}
    \partial_{\zeta}\left\{U\cos{\theta}+(1-\cos{\theta})
    \partial_{\zeta}\phi_2\right\} =0.
\end{equation}
Assuming an integration constant $C_0$ for the above
equation gives
\begin{equation}
    \partial_{\zeta}\phi_2 = \frac{C_0-U\cos{\theta}}
    {1-\cos{\theta}}.
    \label{eq:phi20}
\end{equation}
Varying $\mathcal{L}$ with respect to $\theta$ we find
\begin{align}
    \partial^2_\zeta \theta =
    \sin{\theta}\big\{\cos{\theta} +2U\partial_\zeta
     & \phi_1 - (\partial_\zeta \phi_1)^2
    \nonumber                                     \\
     & -2U\partial_\zeta \phi_2 + (\partial_\zeta
    \phi_2)^2\big\}. \label{eq:theta0}
\end{align}
To avoid divergence of $\partial_{\zeta}^2\theta$ at
infinity, $\partial_\zeta \phi_2$ must be finite at
infinity, which results in the restriction $C_0=U$ and
leads to
\begin{equation}
    \partial_\zeta \phi_2 = U.
    \label{eq:phi2}
\end{equation}
Simplifying Eq.~(\ref{eq:theta0}) with
Eqs.~(\ref{eq:phi1}), (\ref{eq:phi2}) we obtain
\begin{equation}
    \partial^2_\zeta \theta = -U^2\frac{\sin\theta}
    {\cos^4(\theta/2)}+\sin\theta\cos\theta,
    \label{eq:theta}
\end{equation}
whose integration gives the densities
of each component,
\begin{align}
    \frac{n_{1}}{n} & = \frac{1}{2}(1+\cos\theta)
    =
    1-\frac{1-U^2}{1+\abs{U}\cosh(2\sqrt{1-U^2}\zeta)},\nonumber \\
    \frac{n_{2}}{n} & = \frac{1}{2}(1-\cos\theta)
    =
    \frac{1-U^2}{1+\abs{U}\cosh(2\sqrt{1-U^2}\zeta)}.
    \label{eq:density}
\end{align}
Further integrating Eqs.~(\ref{eq:phi1}) and
(\ref{eq:phi2}) gives the phases of both components,
\begin{align}
    \phi_1 & = -\text{sgn}(U)\arctan(\frac{(1-\abs{U})
            \tanh{(\sqrt{1-U^2}\zeta})}{\sqrt{1-U^2}}) + C,
    \nonumber                                          \\
    \label{eq:phase}
    \phi_2 & = U\zeta + \Phi,
\end{align}
where the constant $C$ ensures $\phi_1(\zeta=-\infty)=0$
to fix the $\rm{U}(1)$ gauge. $\Phi$ is a constant
phase shift of the component 2.
Equations~(\ref{eq:density}) and (\ref{eq:phase})
constitute the principle results of this work.

\begin{figure}[htbp]
    \centering
    \includegraphics[width=\linewidth]{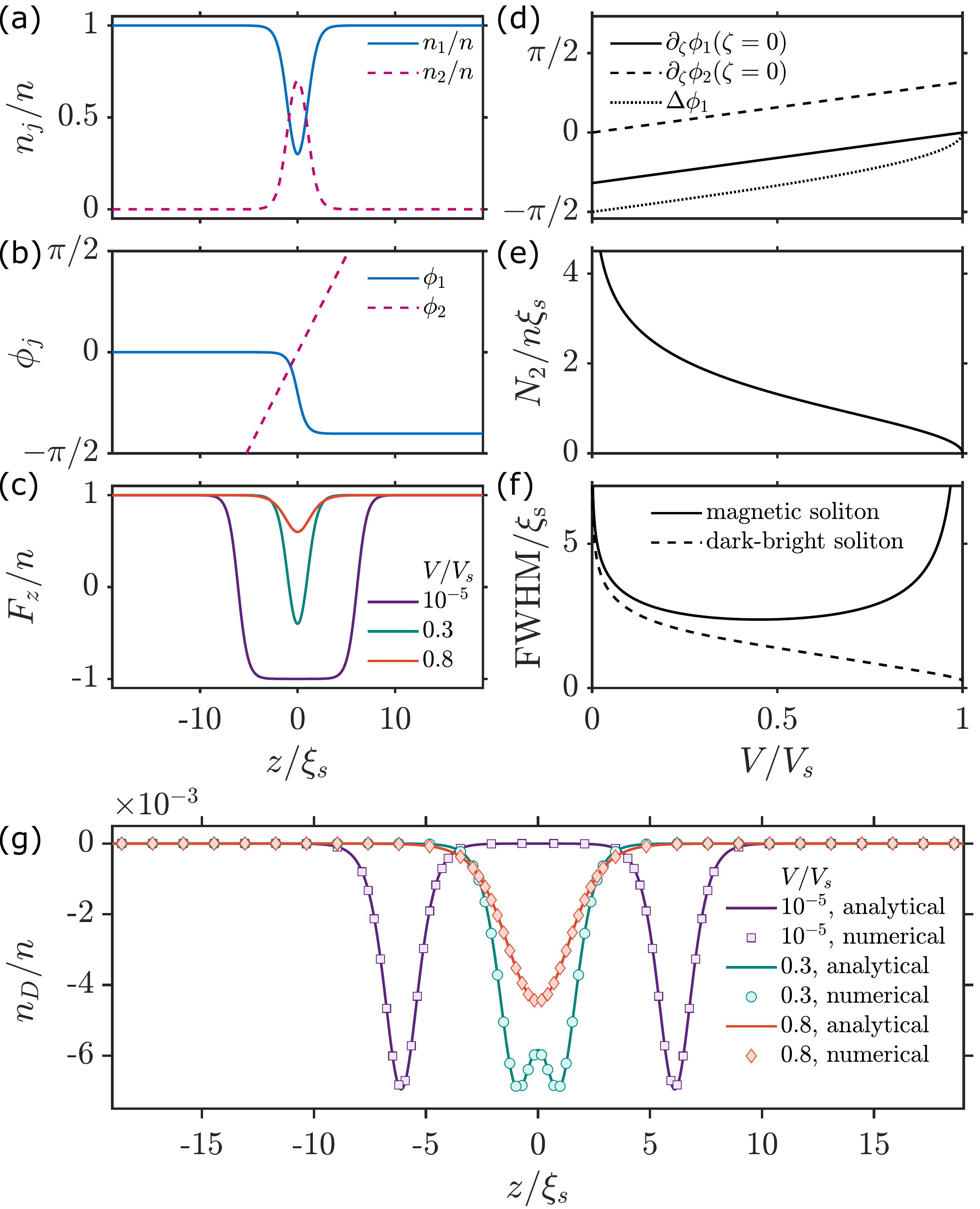}
    \caption{(a,b) Profile of a magnetic soliton with
        $U=0.3$ and $\Phi=0$. The blue solid and red
        dashed curves are (a) density or (b) phase
        profiles of the two components, respectively. (c)
        Spin density of a magnetic soliton with various
        velocities. (d) Dependence of the soliton phases
        on the soliton velocity. The solid and the dashed
        lines are slopes of $\phi_1$, $\phi_2$ at the
        center of the soliton, respectively. The dotted
        curve is the phase jump $\Delta\phi_1 =
        \phi_1(\zeta=+\infty) - \phi_1(\zeta = -\infty)$.
        (e) Total population of the component 2 in a
        magnetic soliton. (f) Full width half maximum
        (FWHM) of the magnetic soliton (dashed line) and
        the dark-bright soliton (solid line). (g) Density
        depletion of the magnetic soliton. The solid
        lines and points represent analytical and
        numerical results, respectively.} \label{fig:sol}
\end{figure}

\section{Soliton Properties}

The soliton solutions (\ref{eq:density}) and
(\ref{eq:phase}) are parametrized by $U$ and $\Phi$. The
phase shift $\Phi$ is only relevant when there exist two
or more solitons, so it will be left aside for now, while
$U = V/V_s$ can take values in $-1 \le U \le 1$. The
maximum speed of the soliton is $V_s=\sqrt{2n\abs{\delta
g}/M}$, which differs from the miscible case by a factor
of two \cite{Qu.Stringari.2016}. Typical density and
phase distributions of a magnetic soliton with immiscible
surrounding condensate are shown in
Figs.~\ref{fig:sol}(a,b), for $U=0.3$ and $\Phi = 0$. The
soliton exhibits a density notch for the component 1,
which is filled by a density bump for the component 2.
The component 2 displays a linear phase with slope
$U/\xi_s$, while the component 1 is featured for its
phase jump $\Delta \phi_1$ across the soliton, which
approaches $\pi/2$ when $U\rightarrow 0$ and vanishes
when $U \rightarrow \pm 1$ (see Fig.~\ref{fig:sol}(d)).
The slope of the phase difference $\partial_z
(\phi_2-\phi_1)$ at the soliton center is
$\mathrm{sgn}(U)/\xi_s$, independent of the speed.

Similar to the magnetic soliton we discuss here, the
dark-bright soliton studied by Busch and Anglin
\cite{Busch.Anglin.2001} comes with a dark component
filled by a bright component, and its phase profiles are
akin to that of magnetic solitons as well. Nevertheless,
significant differences exist in several aspects. First
and most importantly, the dark-bright soliton is
developed under the Manakov regime where $\delta g =0$,
while the immiscible magnetic soliton can only exist when
$\delta g$ is negative. As a consequence, the properties
of a magnetic soliton depend solely on $\delta g$ instead
of $g$. For example, the speed of a dark-bright soliton
is regulated by the sound velocity $c_n=\sqrt{ng/M}$,
while the speed of a magnetic soliton is limited by $V_s
= \sqrt{2n\abs{\delta g}/M}$, which is smaller by
$\sqrt{2\abs{\delta g}/g}\approx 13.6\%$. Here we have
used $\abs{\delta g}/g \approx 0.0093$ for ground state
$\prescript{87}{}{\mathrm{Rb}}$ condensate in
$\ket{F=1,m=\pm1}$, and this ratio will be assumed in the
following discussion.

Secondly, in the low velocity limit the magnetic soliton
exhibits intriguing behaviors unseen in the dark-bright
soliton. As shown in Fig.~\ref{fig:sol}(c), the spin
density (defined as $F_z \equiv n_1 - n_2$) of a magnetic
soliton has a notch. As the velocity approaches zero,
the notch becomes deeper and larger, and eventually it
develops into a pair of magnetic domain walls. Indeed, in
the limit $U\rightarrow 0^+$ the spin density is given by
\begin{align}
    \frac{F_z}{n} \approx 1
    +\tanh(\zeta-\zeta_0/2)
    -\tanh(\zeta+\zeta_0/2),
\end{align}
where $\zeta_0 \xi_s = \xi_s \ln{(2/U)}$ is the
separation between the two domain walls. As $U$ gets
closer to zero, the separation increases significantly
beyond $\xi_s$, the width of the domain walls, and
eventually the background spin is flipped when $U=0$. We
note that the hyperbolic tangent shape of each of these
domain walls is coincident with a recent domain wall
study \cite{Yu.Blakie.2020}.

Thirdly, we consider the bright component population and
the soliton size. Unlike the dark-bright soliton, the
bright component atom number of a magnetic soliton is not
a free parameter, but is dependent on its velocity as
\begin{equation}
    N_2 = n\xi_s\ln({\abs{U}}/(1-\sqrt{1-U^2})).
\end{equation}
As shown in Fig.~\ref{fig:sol}(e), $N_2$ diverges when
$U\rightarrow0$ and vanishes when $U=\pm 1$. Assuming the
dark-bright soliton and the magnetic soliton have the
same bright component population, we compare their full
width half maximum (FWHM) in Fig.~\ref{fig:sol}(f). The
FWHM of a magnetic soliton reaches its minimum value
$2.37\xi_s$ at $U\approx\pm 0.45$ and diverges at
$U\rightarrow\pm1$ or $U\rightarrow 0$, while the FWHM of
a dark-bright soliton monotonically decreases as its
velocity increases.

Finally, we revisit the uniform density approximation.
We find asymptotically \cite{sup} that the density
depletion $n_D$ of a magnetic soliton is given by
\begin{align}
    \frac{n_D}{n}= \frac{\mathfrak{n}-n}{n} \approx \frac{3\delta g}{g}
    \frac{ \abs{U} (1-U^2)
        (\abs{U}+\cosh{(2\sqrt{1-U^2}\zeta)})}{(1+\abs{U}\cosh{(2\sqrt{1-U^2}\zeta}))^2},
    \label{eq:dep}
\end{align}
where $\mathfrak{n}(z,t)$ is reinterpreted as the true
total density distribution of a magnetic soliton and $n =
\lim_{z\rightarrow\infty}\mathfrak{n}(z,t)$ is the
background total density. We compare Eq.~(\ref{eq:dep})
with numerical results obtained from the moving frame
Newton-Raphson method
\cite{sup,Edmonds.Parker.2016,Winiecki.Winiecki.2001,Winiecki.Adams.1999}.
The numerical and analytical results match very well as
illustrated in Fig.~\ref{fig:sol}(g). The density
depletion $n_D/n\sim10^{-3}$ validates the uniform
density approximation. At low soliton velocity, $n_D$
displays a double-dip local core structure with each dip
matching the density depletion of a single magnetic
domain wall as discovered by Yu and Blakie
\cite{Yu.Blakie.2020}. Such a local core feature is
peculiar for the magnetic soliton. In comparison, the
total density of a dark-bright soliton always displays a
dark soliton shape \cite{sup,Busch.Anglin.2001}.

\begin{figure}[htbp] \centering
    \includegraphics[width=\linewidth]{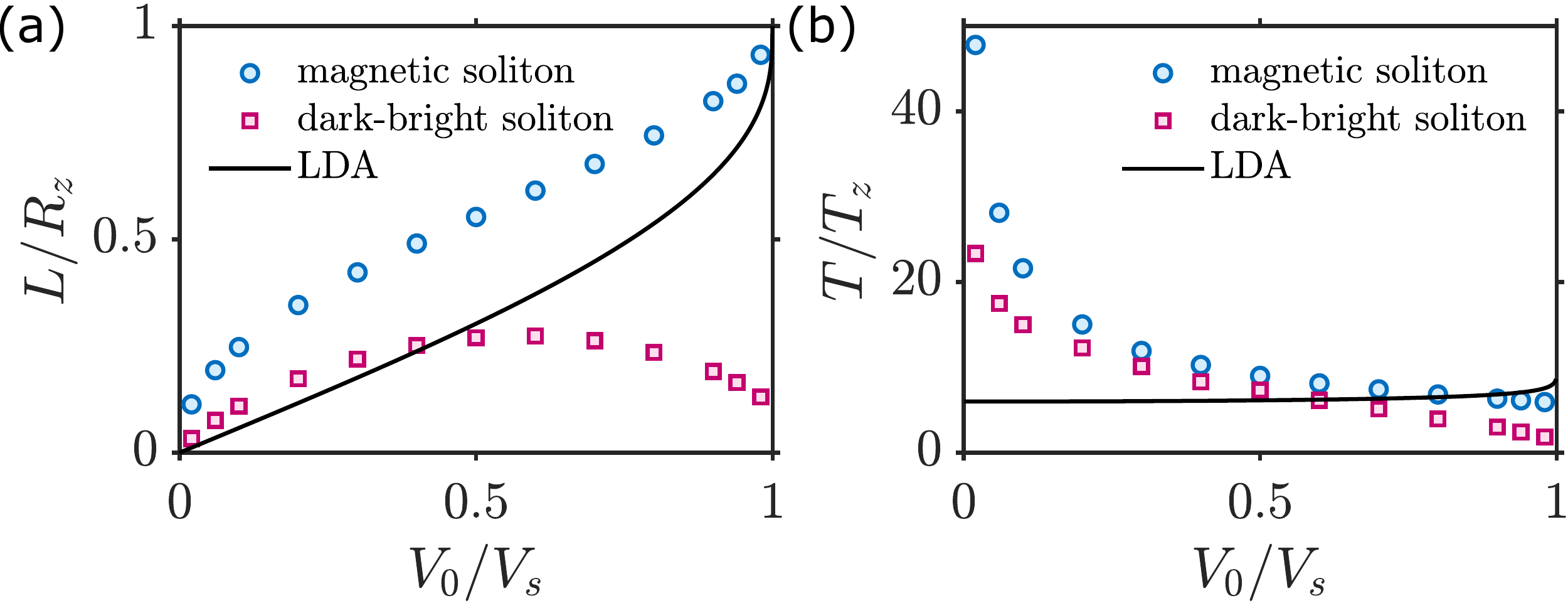}
    \caption{Oscillations of a magnetic soliton and a
    dark-bright soliton in a harmonic trap are compared.
    The blue circles and red squares are numerical
    results for the magnetic soliton and the dark-bright
    soliton, respectively. The black curve is the local
    density approximation (LDA) prediction for the
    magnetic soliton. $V_0$ is the soliton velocity at
    the center of the condensate. In panel (a) we show
    the oscillation amplitude $L$ normalized to the
    Thomas-Fermi radius $R_z=\sqrt{2n_0g/M\omega_z^2}$,
    where $\omega_z$ and $n_0$ are the trapping frequency
    and density at the center. In panel (b) we show
    oscillation period normalized to the trap period
    $T_z=2\pi/\omega_z$.}
    \label{fig:osci}
\end{figure}

\section{Energy and In-trap Oscillation}
The energy of the soliton can be evaluated as the
difference of the total energy $\int \mathcal{E} \dd{z}$
in the presence or absence of the soliton
\cite{pitaevskii2016bose}. Direct calculation gives the
energy $\epsilon=n\hbar V_s\sqrt{1-U^2}$ for a magnetic
soliton in a uniform system, when $U\neq 0$, (when $U=0$,
the energy is zero). The effective mass at small soliton
velocity is $m_\text{eff}=-n\hbar/V_s$, which is
negative, implicating the presence of snake instability
\cite{Nath.Santos.2008}. However, the relatively large
soliton size ($>2.37\xi_s$) establishes marginal
robustness of the solitons against transverse excitations
in a quasi-1D BEC.

The energy of a magnetic soliton in the immiscible regime
exhibits the same form as in the miscible case
\cite{Qu.Stringari.2016}, although in contrast to the
miscible case, the local density approximation (LDA) for
the soliton energy \cite{Konotop.Pitaevskii.2004,sup}
fails to predict the in-trap oscillation of a magnetic
soliton in the immiscible regime we study here (see
Fig.~\ref{fig:osci}). We attribute this discrepancy to
the dependence of $N_2$ on the soliton velocity in the
immiscible case. Both $\epsilon$ and $N_2$ are integrals
of motion of the original Lagrangian (\ref{eq:lag1}) when
$\mathcal{V}$ is non-zero, but the LDA can not
simultaneously guarantee the conservation of these two
quantities when the magnetic soliton oscillates in a trap
with varying velocity. A proper Lagrangian approach
\cite{Kivshar.Krolikowski.1995,Theocharis.Frantzeskakis.2005}
may resolve this problem. For comparison, the oscillation
amplitude and period of a dark-bright soliton are also
displayed in Fig.~\ref{fig:osci}, and the bright
component population is assumed to be the same as that of
the magnetic soliton.

\section{Collision}
\begin{figure}[htbp] \centering
    \includegraphics[width=\linewidth]{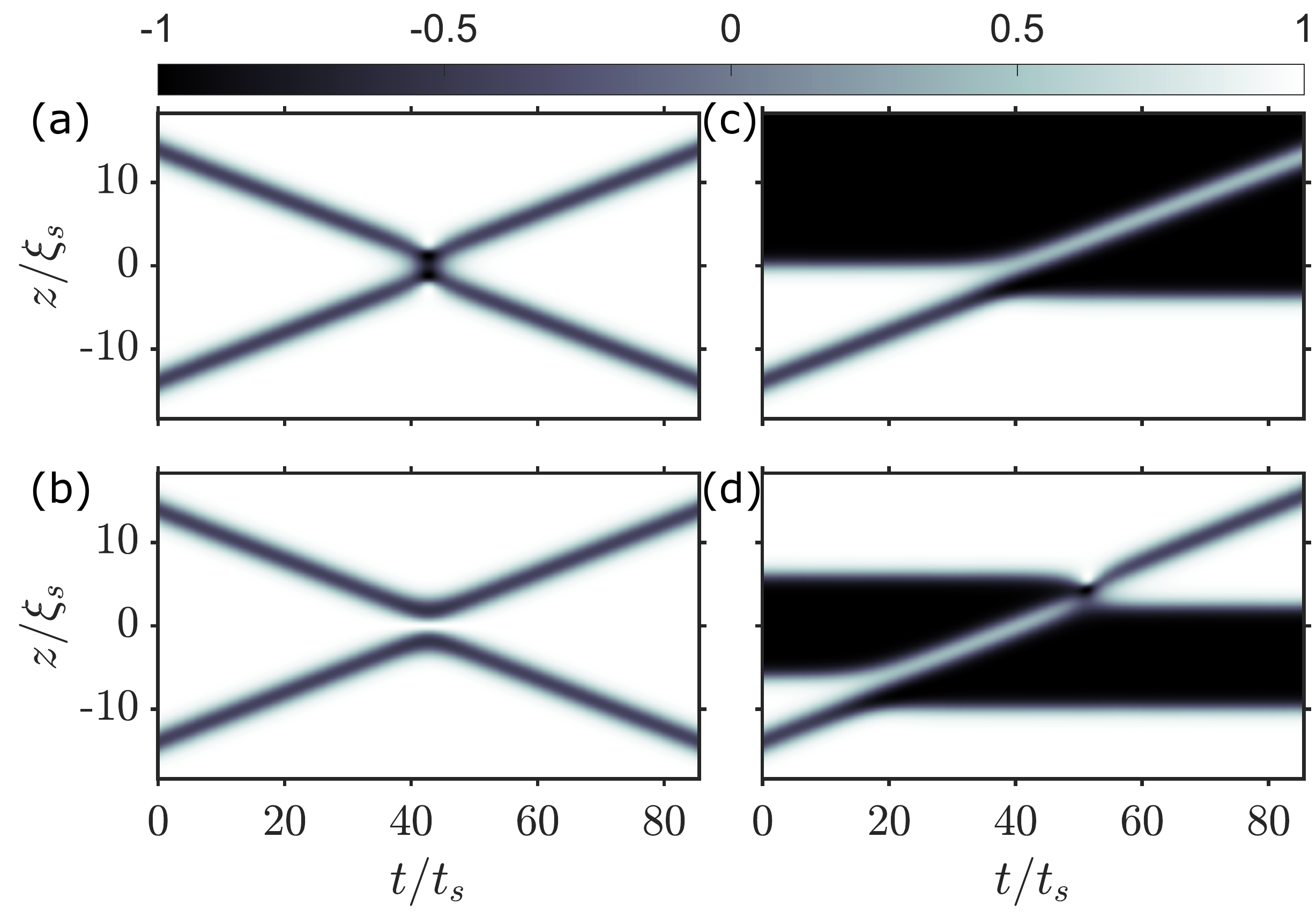}
    \caption{Soliton collisions in a uniform system.
    Plots show the normalized spin density $F_z/n = (n_1
    - n_2)/n$ as a function of space and time. The time
    scale is $t_s = \xi_s/V_s$. In panels (a) and (b),
    $\abs{U}=0.3$ for both solitons. The phase
    differences are (a) $\Delta \Phi = 0$, and (b)
    $\Delta \Phi = \pi$. Panel (c) shows the collision
    between a magnetic soliton with $\abs{U}=0.3$ and a
    static domain wall. Panel (d) shows the collision
    between two magnetic solitons with $\abs{U}=0.3$ and
    $\abs{U}=10^{-5}$.} \label{fig:collision}
\end{figure}

Collisions between two magnetic solitons in an immiscible
BEC depend on their phase $\Phi$. In numerical
simulations, we imprint two magnetic solitons moving
towards each other in a uniform BEC. As shown in
Fig.~\ref{fig:collision}(a), if the phase difference
between the two solitons is zero, i.e., $\Delta \Phi=0$,
the two solitons are found to attract each other during
collision. When $\Delta \Phi = \pi$ the two solitons
repel each other, as illustrated in
Fig.~\ref{fig:collision}(b). Such a behavior is similar
to collisions of dark-bright solitons
\cite{Busch.Anglin.2001}.

Next, we engineer collisions between a magnetic soliton
and a tanh-shaped domain wall \cite{Yu.Blakie.2020}.
Figure~\ref{fig:collision}(c) shows that after collision
the magnetic soliton penetrates the domain wall and its
polarization is flipped. The location of the domain wall
is also shifted after the collision. Collision between a
traveling magnetic soliton and a quasi-static magnetic
soliton (domain wall pair) displays similar dynamics, as
shown in Fig.~\ref{fig:collision}(d), although after
collision the traveling soliton retrieves its initial
shape.

\section{Experimental Generation}
Here we propose a method to experimentally generate a
magnetic soliton in a ferromagnetic spin-1 BEC, where the
two components are taken as the $m=\pm 1$ states. To
eliminate the $m=0$ component, one may introduce a
negative quadratic Zeeman shift $q$, such that the
condensate is forced to stay in the ferromagnetic phase
\cite{Kawaguchi.Ueda.2012}. The length scale of the
soliton is characterized by the spin healing length
$\xi_s$. Using typical experimental conditions for a
quasi-1D $\prescript{87}{}{\text{Rb}}$ BEC
\cite{Bersano.Kevrekidis.2018}, we find the minimum width
of a magnetic soliton is $2.37\xi_s\approx
\SI{9.2}{\micro\meter}$. To avoid snake instability
\cite{Nath.Santos.2008}, the transverse size of the
quasi-1D BEC must be made smaller.

\begin{figure}[htbp] \centering
    \includegraphics[width=\linewidth]{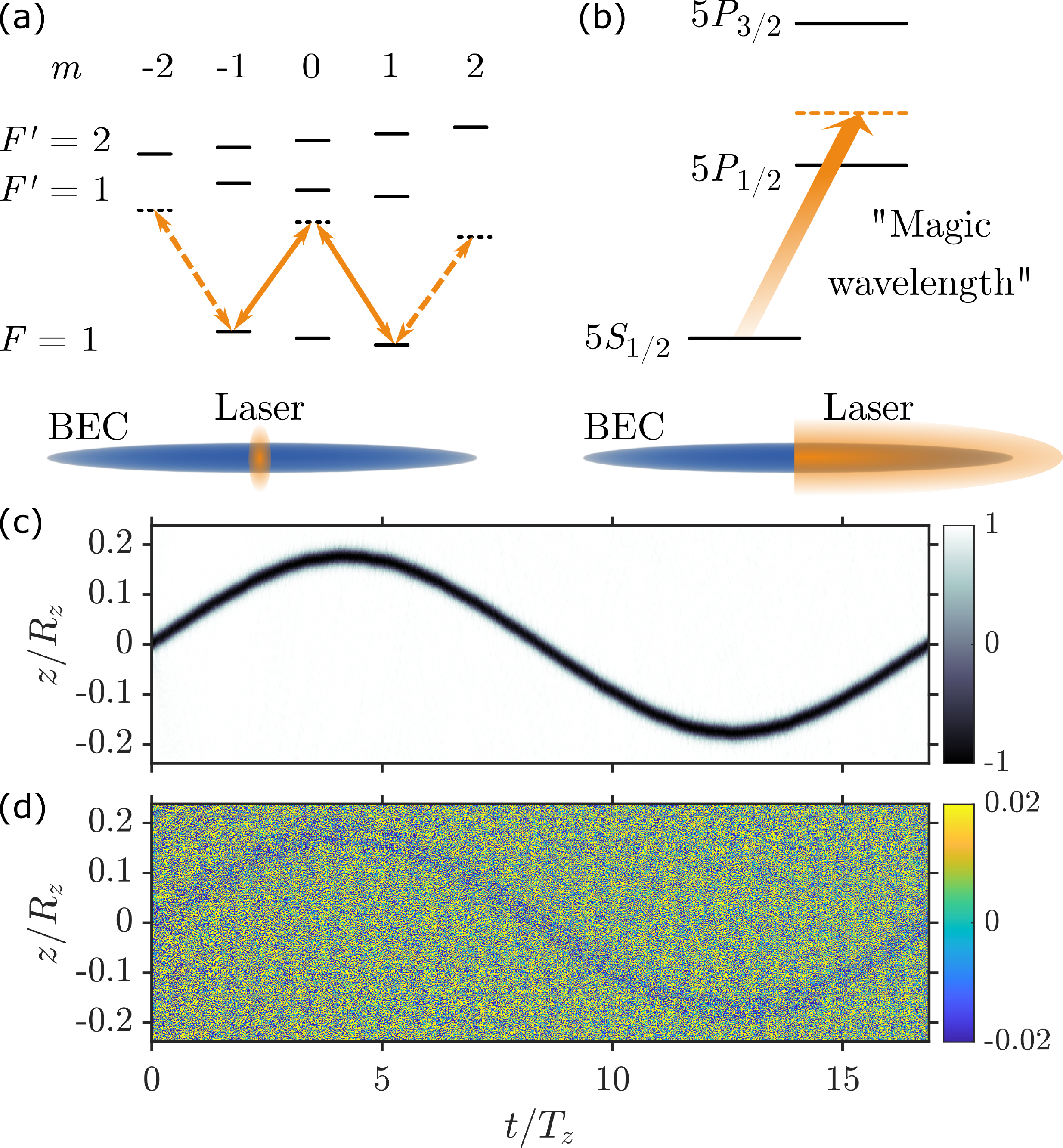}
    \caption{Proposal to generate a magnetic soliton in a
    quasi-1D $\prescript{87}{}{\rm Rb}$ BEC. (a) Local
    population transfer from $m=1$ to $m=-1$. A Raman
    laser pulse coupling the $5S_{1/2},\ket{F=1,m=\pm1}$
    states through the $5P_{1/2}$ state illuminates the
    center of the condensate. The pulse duration is
    controlled to transfer a desired fraction of atoms.
    (b) Magnetic shadow. A enlarged laser beam is imaged
    onto half of the condensate. The laser frequency is
    tuned to the ``magic frequency'' so that it only
    induces vector AC stark shift. The laser beam is
    pulsed such that a finite phase jump is generated.
    (c) Oscillation of the generated magnetic soliton in
    a harmonic trap. Plot shows the normalized spin
    density $ (n_{+1}
    (z,t)-n_{-1}(z,t))/\mathfrak{n}(z,t)$ as a function
    of space and time, where $n_{\pm1}(z,t)$ are the
    densities of the $m=\pm 1$ components. (d) Plot of
    the normalized density depletion defined as
    $(\mathfrak{n}(z,t)-\mathfrak{n}_g(z))/\mathfrak{n}_g(z)$
    where $\mathfrak{n}_g(z)$ is the ground state density
    distribution.} \label{fig:generation}
\end{figure}

Suppose initially the condensate is prepared in a
ferromagnetic state with all the atoms in the $m=1$ state
and stabilized by a negative quadratic Zeeman shift. To
generate a magnetic soliton we first apply a local
population transfer from $m=1$ to $m=-1$, which can be
accomplished by a focused Raman laser pulse
\cite{Wright.Bigelow.2008}, as shown in
Fig.~\ref{fig:generation}(a). Subsequently a magnetic
shadow \cite{Chai.Raman.2020} (see
Fig.~\ref{fig:generation}(b)) is cast to induce a phase
difference, leading to a local relative superfluid
velocity between the two components. The relative
superfluid velocity then helps to assist in the formation
of a magnetic soliton.

The above procedure is confirmed in numerical simulation
\cite{sup} and indeed a single magnetic soliton is
generated which subsequently oscillates in a harmonic
trap, as shown in Fig.~\ref{fig:generation}(c). To be
more realistic, we included Gaussian noise and a
negative quadratic Zeeman shift in our simulation.
The density depletion of the generated
soliton, shown in Fig.~\ref{fig:generation}(d), displays
a double-dip core structure, which is a characteristic
feature of the magnetic soliton. The fringes in
Fig.~\ref{fig:generation}(d) are density waves as
byproducts of our procedure.

\section{Conclusion and Outlook}
We have derived a closed-form magnetic soliton solution
for the coupled two-component Gross-Pitaevskii equations
with $\delta g<0$. We hope our results will stimulate
experimental studies. Though the solution is obtained in
a two-component system, it can be extended to a broader
class of soliton solutions in a spin-1 system by
exploiting the underline SO(3) symmetry
\cite{Chai.Raman.2020jrk,Yu.Blakie.2020}. The correlation
between the quench dynamics of a ferromagnetic spin-1
condensate \cite{Fujimoto.Ueda.2018} and magnetic
solitons is an interesting topic worthy of some immediate
studies. Other unsolved problems, including the dynamical
stability in higher dimensions and the in-trap
oscillation of a magnetic soliton, remain to be explored
in the future.

\section{Acknowledgement}
We thank fruitful discussion with Kazuya Fujimoto and Di
Lao. X.C. and C.R. acknowledge the support from the
National Science Foundation through award no. 2011478.
L.Y. acknowledges the support from the National Key R\&D
Program of China (Grants No. 2018YFA0306504), and from
the National Natural Science Foundation of China (NSFC)
(No. 11654001, No. 91736311, No. 91836302, and No.
U1930201).
\bibliographystyle{apsrev4-1}
\bibliography{Magsol_F.bib}

\section{Supplementary Material}
\beginsupplement
\subsection{Spin-1 Gross-Pitaevskii equations in 3D}
A spin-1 BEC can be well described by a spinor wave
function $\Psi_m(\bm{r},t)$, where $m=-1,0,+1$ is the
magnetic quantum number and $\bm{r},~t$ are space and
time coordinates, respectively. The dynamics of
$\Psi_m(\bm{r},t)$ is governed by three coupled
Gross-Pitaevskii equations (GPEs),
\begin{align}
    i\hbar \pdv{t}
    \Psi_{m}=(-\frac{\hbar^2}{2M}\grad^2 & +\mathcal{V})\Psi_m
    + qm^2\Psi_m + c_0 \mathfrak{n} \Psi_m \nonumber
    \\
                                         & + c_2 \sum_{n=-1}^{1} \bm{F} \cdot
    (\hat{\bm{F}})_{mn}\Psi_n,
    \label{eq:gp3D}
\end{align}
where $M$ is the atomic mass. $\mathcal{V}(\bm{r}),~q$
are the spin independent potential and the quadratic
Zeeman shift, respectively.
$\mathfrak{n}(\bm{r},t)=\sum_{m=-1}^1
\abs{\Psi_m(\bm{r},t)}^2$ is the total density. The wave
function is normalized to the total number of atoms as
$\int \dd{\bm{r}} \mathfrak{n} (\bm{r},t) = N$.
$c_0,~c_2$ are spin independent and spin dependent
interaction coupling constants defined as $c_0 = 4 \pi
\hbar^2 (a_2+2a_0)/3M$ and $c_2 = 4\pi\hbar^2(a_2 -
a_0)/3M$, where $a_0,~a_2$ are $s$-wave scattering
lengths of collisions in the total $F=0,2$ channels. We
consider ferromagnetic interaction only in this study
such that $c_2<0<c_0$. $\bm{F} (\bm{r},t)=\sum_{m,n=-1}^1
\Psi_m^*(\bm{r},t)(\hat{\bm{F}})_{mn}\Psi_n(\bm{r},t)$ is
the spin density and $\hat{\bm{F}} =
(\hat{F}_x,\hat{F}_y,\hat{F}_z)^T$ with
$\hat{F}_x,~\hat{F}_y,~\hat{F}_z$ being the spin-1
matrices,
\begin{align}
    \hat{F}_x & =
    \frac{1}{\sqrt{2}}
    \begin{pmatrix}
        0 & 1 & 0 \\
        1 & 0 & 1 \\
        0 & 1 & 0
    \end{pmatrix}
    ,~
    \hat{F}_y =
    \frac{i}{\sqrt{2}}
    \begin{pmatrix}
        0 & -1 & 0  \\
        1 & 0  & -1 \\
        0 & 1  & 0
    \end{pmatrix}
    ,\nonumber    \\
    \hat{F}_z & =
    \begin{pmatrix}
        1 & 0 & 0  \\
        0 & 0 & 0  \\
        0 & 0 & -1
    \end{pmatrix}.
\end{align}
In experiments, BECs are usually trapped optically
and the trapping potential $\mathcal{V}(\bm{r})$ can be
approximated as a harmonic potential. We use experimental
parameters from Ref.~\cite{Bersano.Kevrekidis.2018} where
the cigar-shaped trap has frequencies
$\{\omega_x,\omega_y,\omega_z\} = 2\pi \times
    \{176,174,1.4\}~\si{\hertz}$ (we have changed the labels
for consistency with our paper). With $N=0.8 \times 10^6$
atoms in the BEC, the Thomas-Fermi radii are
$\{R_x,R_y,R_z\} = {\{3,3,369\}}~\si{\micro\meter}$.

\subsection{Spin-1 Gross-Pitaevskii equations in 1D}
For a cigar-shaped condensate with $\omega_x \gg \omega_
    z$ and $\omega_y \gg \omega_z$, one can assume that the
wave function can be written as
\begin{equation}
    \Psi_m(\bm{r},t) = \Psi^{(\mathrm{1D})}_m(z,t) G(x,y),
\end{equation}
where $G(x,y)$ is the transverse wavefunction in the
Thomas-Fermi limit:
\begin{eqnarray}
    G(x,y) =
    \begin{cases}
        \displaystyle \sqrt{\frac{2}{\pi R_x R_y}
            \left( 1 - \frac{x^2}{R_x^2} - \frac{y^2}{R_y^2}
        \right)}, & (\frac{x^2}{R_x^2}+\frac{y^2}{R_y^2}\leq 1
        );                                                    \\ \displaystyle 0, &
        ({\rm otherwise}).
    \end{cases}
    \nonumber \\
\end{eqnarray}
$G(x,y)$ and $\Psi^{(\mathrm{1D})}_m(z,t)$ are normalized
independently as $\int \dd{x}\dd{y} \abs{G(x,y)}^2 = 1$
and $\int \dd{z} \sum_{m=-1}^1 \abs{\Psi^{(\mathrm{1D})}_m(z,t)}^2 = N$.
The 3D GPEs (\ref{eq:gp3D}) can then be reduced to
\begin{align}
    i\hbar \pdv{t}
     & \Psi^{(\mathrm{1D})}_{m}=(-\frac{\hbar^2}{2M}\pdv[2]{}{z}+\mathcal{V}^{\rm (1D)})\Psi^{(\mathrm{1D})}_m
    + qm^2\Psi^{(\mathrm{1D})}_m \nonumber                                                                     \\
     & + g_0 \mathfrak{n}^{\rm (1D)}
    \Psi^{(\mathrm{1D})}_m + g_2 \sum_{n=-1}^{1}
    \bm{F}^{\rm (1D)} \cdot
    (\hat{\bm{F}})_{mn}\Psi^{(\mathrm{1D})}_n,
    \label{eq:gp1D}
\end{align}
where $g_0,~g_2$ are effective coupling constants in 1D
given by $g_0 = 4c_0/3\pi R_x R_y$ and $g_2 = 4c_2/3\pi
    R_x R_y$. The definitions for total density and spin
density in 1D are given accordingly as
$\mathfrak{n}^{\rm (1D)}(z,t)=\sum_{m=-1}^1
    \abs{\Psi^{(\mathrm{1D})}_m(z,t)}^2$ and $\bm{F}^{\rm (1D)}(z,t)
    =\sum_{m,n=-1}^1
    \Psi^{(\mathrm{1D})*}_m(z,t)(\hat{\bm{F}})_{mn}\Psi^{(\mathrm{1D})}_n(z,t)$.
$\mathcal{V}^{\rm (1D)}(z)=M\omega_z^2 z^2/2$ is the
spin-independent potential in the presence of a harmonic
trap.

\subsection{Binary Gross-Pitaevskii equations in 1D}
Experimentally one can use microwave dressing to apply a
negative quadratic Zeeman shift. Hence the energy of
$m=\pm 1$ states is lowered so that the spin exchange
collision $\ket{1,1} + \ket{1,-1} \rightarrow 2\ket{1,0}$
can be suppressed. With $m=0$ atoms eliminated, the
spin-1 GPEs~(\ref{eq:gp1D}) reduce to the binary GPEs,
\begin{align}
    i \hbar \frac{\partial}{\partial t}\psi_1 & = \left (
    -\frac{\hbar^2}{2M} \pdv[2]{z} + \mathcal{V}^{\rm (1D)} + g_{11}|\psi_1|^2 +
    g_{12}|\psi_{2}|^2 \right ) \psi_1, \nonumber         \\
    i \hbar
    \frac{\partial}{\partial t}\psi_{2}       & = \left (
    -\frac{\hbar^2}{2M} \pdv[2]{z} + \mathcal{V}^{\rm (1D)} + g_{22}|\psi_{2}|^2 +
    g_{12}|\psi_{1}|^2 \right ) \psi_{2},
    \label{eq:bGP}
\end{align}
where $\psi_1 \equiv \Psi_{+1} ^ {\rm (1D)}$ and $\psi_2
    \equiv \Psi_{-1} ^ {\rm (1D)}$. $g_{11} = g_{22} = g_0 +
    g_2 = g$ are the intraspecies interaction strengths. $g_{12}
    = g_0 - g_2 = g - \delta g$ is the interspecies interaction strengths.
The quadratic Zeeman
shift term has been eliminated because it only introduces
a constant energy shift for the two states $m=\pm 1$.
Equations.~(\ref{eq:bGP}) can be derived from the
Lagrangian (\ref{eq:lag1}) given in the main text,
provided that the label (1D) is removed.

\subsection{Dimensionless spin-1 GPEs in 1D}
We choose $z_0 = \sqrt{\hbar/\omega_z M}$, $t_0 =
1/\omega_z$, and $\epsilon_0 = \hbar \omega_z$ as our
length, time, and energy scales, respectively. Then the
dimensionless spin-1 GPEs are written as
\begin{align}
    i\pdv{\tilde{t}}
    \tilde{\Psi}_{m}= & (-\frac{1}{2}\pdv[2]{}{\tilde{z}}+\tilde{\mathcal{V}})\tilde{\Psi}_m
    + \tilde{q} m^2 \tilde{\Psi}_m \nonumber                                                 \\
                      & + \tilde{g}_0 \tilde{\mathfrak{n}}
    \tilde{\Psi}_m + \tilde{g}_2 \sum_{n=-1}^{1}
    \tilde{\bm{F}} \cdot
    (\hat{\bm{F}})_{mn}\tilde{\Psi}_n,
    \label{eq:gpdl}
\end{align}
where the dimensionless quantities are given in the
following,
\begin{gather}
    \tilde{z} = \frac{z}{z_0},~
    \tilde{t} = \frac{t}{t_0},~
    \tilde{\mathcal{V}} =
    \frac{\mathcal{V}^{\rm (1D)}}{\epsilon_0},~ \tilde{q}
    = \frac{q}{\epsilon_0},
    \nonumber \\
    \tilde{g}_0 = \frac{g_0 N}{x_0\epsilon_0},~
    \tilde{g}_2 = \frac{g_2 N}{x_0\epsilon_0},
    \nonumber \\
    \tilde{\Psi}_m =
    \sqrt{\frac{x_0}{N}}\Psi_m^{\rm (1D)},~
    \tilde{\mathfrak{n}} = \frac{x_0}{N}\mathfrak{n}^{(\rm 1D)},~
    \tilde{\bm{F}} = \frac{x_0}{N}\bm{F}^{(\rm 1D)}.
\end{gather}
The dimensionless wavefunction is normalized as
$\int \dd{\tilde{z}} \sum_{m=-1}^1
    \abs{\tilde{\Psi}_m(\tilde{z},\tilde{t})}^2 = 1$. Using
typical experimental parameters in
Ref.~\cite{Bersano.Kevrekidis.2018} and scattering
lengths data in Ref.~\cite{Kempen.Verhaar.2002}, we find
the nonlinear coefficients are $\tilde{g}_0 = 23729$ and
$\tilde{g}_2 = -110$.

\subsection{Newton-Raphson method}
Here we discuss how we numerically obtain the true
magnetic soliton solutions. Consider a stationary
solution solution of the GPEs
(\ref{eq:gpdl}),
\begin{equation}
    \tilde{\Psi}_m (\tilde{z},\tilde{t}) = \tilde{\Psi}_m
    (\tilde{z}) e^{-i \tilde{\mu} \tilde{t}},
    \label{eq:ss}
\end{equation}
where $\tilde{\mu}$ is the dimensionless chemical
potential. Substituting Eq.~(\ref{eq:ss}) back into
Eq.~(\ref{eq:gpdl}), we have the time-independent GPEs,
\begin{align}
    \tilde{\mu}\tilde{\Psi}_{m}= & (-\frac{1}{2}\pdv[2]{}{\tilde{z}}+\tilde{\mathcal{V}})\tilde{\Psi}_m
    + \tilde{q} m^2 \tilde{\Psi}_m \nonumber                                                            \\
                                 & + \tilde{g}_0 \tilde{n}_{\mathrm{tot}}
    \tilde{\Psi}_m + \tilde{g}_2 \sum_{n=-1}^{1}
    \tilde{\bm{F}} \cdot
    (\hat{\bm{F}})_{mn}\tilde{\Psi}_n.
    \label{eq:gpti}
\end{align}
Since we are interested in traveling solitons in a
uniform system, we assume $\tilde{\mathcal{V}} =
    \tilde{q}= 0$ and switch to the moving frame with
velocity $\tilde{V}$, where the moving-frame
time-independent GPEs \cite{Winiecki.Winiecki.2001} takes
the forms
\begin{align}
    \tilde{\mu}\tilde{\Psi}_{m}=(-\frac{1}{2}\pdv[2]{}{\tilde{z}}
    + i\tilde{V} \pdv{}{\tilde{z}} ) & \tilde{\Psi}_m
    + \tilde{g}_0 \tilde{n}_{\mathrm{tot}}
    \tilde{\Psi}_m
    \nonumber                                                      \\
    +                                & \tilde{g}_2 \sum_{n=-1}^{1}
    \tilde{\bm{F}} \cdot
    (\hat{\bm{F}})_{mn}\tilde{\Psi}_n.
    \label{eq:gpti2}
\end{align}
To numerically find stationary magnetic soliton solutions
of Eq.~(\ref{eq:gpti2}) we use the Newton-Raphson method
which has been used to obtain dipolar solitons or vortex
in moving-frame
\cite{Edmonds.Parker.2016,Winiecki.Winiecki.2001,Winiecki.Adams.1999}.
The simulation is performed on a 1D line $\tilde{z} \in
    [-40,40]$ discretized into $\mathcal{N} = 4096$ girds
with spacing $\Delta \tilde{z} = 80/(\mathcal{N}-1)$. The
discretized wavefunction is descried by 
$\tilde{\Psi}_{j,m}$ where $j=1,2,...,\mathcal{N}$
denotes the $j$-th grid and $m=0,\pm1$ is the magnetic
quantum number. Since the real and imaginary parts of the
wavefunction are independent degrees of freedom, we
define $\tilde{\Psi}_{j,r,m}$ with $r=0,1$, where
$\tilde{\Psi}_{j,0,m} = \mathrm{Re}(\tilde{\Psi}_{j,m})$
and $\tilde{\Psi}_{j,1,m} =
    \mathrm{Im}(\tilde{\Psi}_{j,m})$. Eq.~(\ref{eq:gpti2})
can then be discretized as $\bm{f}(\tilde{\bm{\Psi}}) =
    0$ where
\begin{align}
    f_{j,r,m} = & -\frac{1}{2}\frac{\tilde{\Psi}_{j-1,r,m}
        - 2\tilde{\Psi}_{j,r,m} + \tilde{\Psi}_{j+1,r,m}}{(\Delta \tilde{z})^2}
    \nonumber                                                    \\
                & +(2r-1)\tilde{V}\frac{\tilde{\Psi}_{j+1,1-r,m}
        - \tilde{\Psi}_{j-1,1-r,m}}{2\Delta \tilde{z}}
    \nonumber                                                    \\
                & + (- \tilde{\mu} + \tilde{g}_0 \sum_{m',r'}
    \tilde{\Psi}_{j,r',m'}^2 ) \tilde{\Psi}_{j,r,m}
    \nonumber                                                    \\
                & + \tilde{g}_2 \sum_{m'}
    (\tilde{F}_{x,j} \hat{F}_{x,mm'} + \tilde{F}_{z,j}
    \hat{F}_{z,mm'}) \tilde{\Psi}_{j,r,m'}
    \nonumber                                                    \\
                & - i(2r-1)\tilde{g}_2 \sum_{m'} \tilde{F}_{y,j}
    \hat{F}_{y,mm'} \tilde{\Psi}_{j,1-r,m'},
\end{align}
and where $\tilde{\bm{F}}_{j}$ is the discretized spin
density evaluated at the $j$-th grid,
\begin{align}
    \tilde{\bm{F}}_j & = \sum_{m',n'=-1}^1
    \tilde{\Psi}_{j,m'}^*(\hat{\bm{F}})_{m'n'}\tilde{\Psi}_{j,n'} \nonumber \\
                     & = \sum_{m',n'=-1}^1
    (\tilde{\Psi}_{j,0,m'}-i\tilde{\Psi}_{j,1,m'})(\hat{\bm{F}})_{m'n'}
    (\tilde{\Psi}_{j,0,n'}+i\tilde{\Psi}_{j,1,n'}).
\end{align}
We impose the Neumann boundary condition such that at the
factitious grids $j=0$ and $j = \mathcal{N}+1$ the
wavefunctions are given by
\begin{equation}
    \tilde{\Psi}_{0,r,m} = \tilde{\Psi}_{2,r,m},~
    \tilde{\Psi}_{\mathcal{N}+1,r,m}
    = \tilde{\Psi}_{\mathcal{N}-1,r,m}.
\end{equation}
Starting from the analytical wavefunction of a magnetic
soliton given in Eqs.~(\ref{eq:density}) and (\ref{eq:phase}), Newton-Raphson method
solves $\bm{J}\delta \tilde{\bm{\Psi}} = - \bm{f}$ for
$\delta\tilde{\bm{\Psi}}$ to update the wavefunction as
$\tilde{\bm{\Psi}}_{p+1} = \tilde{\bm{\Psi}}_{p} +
    \delta\tilde{\bm{\Psi}}$ at each step $p$, where $\bm{J}$
is the Jacobian of $\bm{f}$ with respect to
$\tilde{\bm{\Psi}}$,
\begin{align}
    J_{\substack{{j,r,m}                                                    \\ {k,s,n}}} = &
    \pdv{f_{j,r,m}}{\tilde{\Psi}_{k,s,n}} \nonumber                         \\
    = & -\frac{1}{2} \frac{\delta_{j+1,k} -
        2\delta_{j,k}+ \delta_{j-1,k}}{(\Delta \tilde{z})^2}
    \delta_{r,s}\delta_{m,n} \nonumber                                      \\
      & + (2r-1) \tilde{V}
    \frac{\delta_{j+1,k}-\delta_{j-1,k}}{2\Delta \tilde{z}}
    \delta_{1-r,s}\delta_{m,n} \nonumber                                    \\
      & + (-\tilde{\mu} + \tilde{g}_0 \sum_{m',r'}
    \tilde{\Psi}_{j,r',m'}^2)\delta_{j,k}\delta_{r,s}\delta_{m,n} \nonumber \\
      & + 2\tilde{g}_0 \tilde{\Psi}_{j,r,m}
    \tilde{\Psi}_{j,s,n} \delta_{j,k}  \nonumber                            \\
      & + \tilde{g}_2
    (\tilde{F}_{x,j} \hat{F}_{x,mn} + \tilde{F}_{z,j}
    \hat{F}_{z,mn})
    \delta_{j,k}\delta_{r,s}
    \nonumber                                                               \\
      & + \tilde{g}_2 \sum_{m'}
    (K_{x,j,s,n} \hat{F}_{x,mm'} \nonumber                                  \\
      & \hspace{2cm}+ K_{z,j,s,n}
    \hat{F}_{z,mm'}) \tilde{\Psi}_{j,r,m'} \delta_{j,k}
    \nonumber                                                               \\
      & - i(2r-1)\tilde{g}_2 \tilde{F}_{y,j}
    \hat{F}_{y,mn} \delta_{j,k}\delta_{1-r,s}
    \nonumber                                                               \\
      & - i(2r-1)\tilde{g}_2 \sum_{m'} K_{y,j,s,n}
    \hat{F}_{y,mn} \tilde{\Psi}_{j,1-r,m'}\delta_{j,k},
\end{align}
and where
\begin{align}
    \bm{K}_{j,s,n} = & \sum_{n'}
    \biggl\{
    (\hat{\bm{F}})_{n,n'} + (\hat{\bm{F}})_{n,n'}^*)
    \tilde{\Psi}_{j,s,n'}
    \nonumber                    \\
                     & -i(2s-1)
    ((\hat{\bm{F}})_{n,n'} - (\hat{\bm{F}})_{n,n'}^*)
    \tilde{\Psi}_{j,1-s,n'}\biggr\}.
\end{align}
Since the atom number is fixed in our simulation, at each step we update the chemical potential $\tilde{\mu}$ according to Eq.~(\ref{eq:gpti2}). Such iteration can converge at a final wavefunction
$\tilde{\bm{\Psi}}_f$ satisfying
$\bm{f}(\tilde{\bm{\Psi}}_f) = 0$, which is the true
magnetic soliton solution we seek to obtain. The
convergence is determined once the correction
$\abs{\delta\tilde{\bm{\Psi}}}$ is smaller than an
arbitrary tolerance.

\subsection{Uniform density approximation and asymptotic
    form of the density depletion} Consider a parametrization
for the condensate wave functions beyond the uniform
density approximation:
\begin{equation}
    \mqty(\psi_1\\\psi_2) = \sqrt{\mathfrak{n}}
    \mqty(\cos{(\theta/2)}e^{i\phi_1}\\
    \sin{(\theta/2)}e^{i\phi_2})
    e^{-i \mu t / \hbar},
\end{equation}
where $\mu=ng$ is the chemical potential at equilibrium
with only one component present at density $n$.
$\mathfrak{n}(z,t)$ is the total density as a function of
space and time. Using dimensionless variables
$\mathcal{Z} = z/\xi_s$ and $\mathcal{T} = t/t_s$, we
find the Lagrangian is given by
\begin{align}
    \frac{\mathcal{L}}{n M V_s^2}
    = & \frac{g}{4n^2 \delta
        g }  (\mathfrak{n}^2 - 2 n
    \mathfrak{n})\nonumber              \\
      & -\frac{1}{8n^2}
    \biggl\{
    \mathfrak{n}^2 \sin^2\theta +
    \frac{n(\partial_{\mathcal{Z}}\mathfrak{n})^2}{\mathfrak{n}}
    +n \mathfrak{n} (\partial_{\mathcal{Z}}\theta)^2
    \nonumber                           \\
      & \hspace{1.2cm}+ 4n \mathfrak{n}
    (1+\cos{\theta})\partial_\mathcal{T} \phi_1
    \nonumber                           \\
      & \hspace{1.2cm}+ 4n \mathfrak{n}
    (1-\cos{\theta})\partial_\mathcal{T} \phi_2
    \nonumber                           \\
      & \hspace{1.2cm}+ 2n \mathfrak{n}
    (1+\cos{\theta})(\partial_\mathcal{Z} \phi_1)^2
    \nonumber                           \\
      & \hspace{1.2cm}+2n \mathfrak{n}
    (1-\cos{\theta})(\partial_\mathcal{Z} \phi_2)^2
    \bigg\},
\end{align}
in the absence of trapping potential. Variation of the
Lagrangian with respect to
$\mathfrak{n}$ gives
\begin{align}
    \frac{\mathfrak{n}-n}{n} = & \frac{\delta g}{2g}
    \biggl\{
    \frac{\mathfrak{n} \sin^2\theta}{n} +
    \frac{(\partial_{\mathcal{Z}}\mathfrak{n})^2}{2\mathfrak{n}^2}
    -\frac{\partial^2_{\mathcal{Z}}\mathfrak{n}}{\mathfrak{n}}
    + \frac{1}{2} (\partial_{\mathcal{Z}}\theta)^2
    \nonumber                                        \\
                               & \hspace{0.5cm}+ 2
    (1+\cos{\theta})\partial_\mathcal{T} \phi_1
    + 2
    (1-\cos{\theta})\partial_\mathcal{T} \phi_2
    \nonumber                                        \\
                               & \hspace{0.5cm}+
    (1+\cos{\theta})(\partial_\mathcal{Z} \phi_1)^2
    +
    (1-\cos{\theta})(\partial_\mathcal{Z} \phi_2)^2
    \bigg\}.
\end{align}
The right-hand side (RHS) of the above equation becomes
negligible when $\abs{\delta
        g}/g\ll 1$ such that $\mathfrak{n}\approx n$ at the
leading order, which validates our uniform density
approximation. Inserting the magnetic soliton solution
obtained in the main text into the RHS of the above
equation, we have the asymptotic density depletion,
\begin{align}
    \frac{n_D}{n}= \frac{\mathfrak{n}-n}{n} \approx \frac{3\delta g}{g}
    \frac{ \abs{U} (1-U^2)
        (\abs{U}+\cosh{(2\sqrt{1-U^2}\zeta)})}{(1+\abs{U}\cosh{(2\sqrt{1-U^2}\zeta}))^2},
\end{align}
where $\zeta = \mathcal{Z} - U \mathcal{T}$. Then the
total population depletion becomes
\begin{equation}
    N_D = \int \dd{\zeta} n_D\xi_s \approx
    \frac{3n\delta g}{g} \sqrt{1-U^2} \xi_s.
\end{equation}

\subsection{Comparison with dark-bright soliton}
If we assume $\delta g=0$ then Eqs.~(\ref{eq:bGP}) reduce
to a two-component Manakov system, which has a
dark-bright soliton solution \cite{Busch.Anglin.2001},
\begin{align}
    \psi_1 & = \sqrt{n} \cos{\alpha}
    \tanh{\left\{\kappa(z-kt)\right\}} + i\sqrt{n}
    \sin{\alpha},                                                           \\
    \psi_2 & = \sqrt{\frac{N_B \kappa}{2}} \sech{\left[\kappa(z-kt)\right]}
    \exp{i \kappa z \tan{\alpha} + i\Omega_B t},
\end{align}
where
\begin{align}
    \kappa   & = \frac{1}{\xi_n}\left\{\sqrt{\cos^2{\alpha}+\left(\frac{N_B}{4n\xi_n} \right)^2} - \frac{N_B}{4n\xi_n} \right\}, \\
    k        & = \frac{\hbar}{M} \kappa \tan{\alpha},                                                                           \\
    \Omega_B & = \frac{\hbar}{M}\kappa^2(1-\tan^2\alpha)/2,
\end{align}
and $\xi_n = \hbar/\sqrt{n M g}$ is the healing
length. The solution is controlled by two parameters: the
total particle number of the bright component $N_B$ and
the velocity angle $\alpha$. The soliton velocity is $k$,
whose maximum value is the sound velocity $c_n =
    \sqrt{ng/M}$. The full width half maximum of the bright
component is $2\rm{arccosh}(\sqrt{2})/\kappa$. The total
density is
\begin{equation}
    \mathfrak{n} = n-n\kappa^2 \xi_n^2 \sech^2\{\kappa (z-kt)\},
\end{equation}
and the total depletion is
\begin{equation}
    N_D = -2n\kappa\xi_n^2.
\end{equation}
According to \cite{Busch.Anglin.2001}, the low-velocity
equation of motion for a dark-bright soliton in an
inhomogeneous potential is given by
\begin{align}
    \ddot{q} = -\frac{\mathcal{V}'(q)}{2M}\left(1- \frac{
        N_B/4n\xi_n}{\sqrt{-\mathcal{V}(q)/Mc_n^2+(N_B/4n\xi_n)^2+1}}\right).
\end{align}
\subsection{Oscillation of a magnetic soliton or a
    dark-bright soliton in a harmonic trap}
To obtain the oscillation period and amplitude of a
magnetic soliton or a dark-bright soliton, we numerically
solve Eq.~(\ref{eq:gpdl}) with a time step $\Delta
    \tilde{t}=1.885\times10^{-4}$. For 
oscillation of a dark-bright soliton, we set
$\tilde{g}_2=0$. We first find the ground state of the
condensate $\tilde{\psi}_g(\tilde{z})$ by propagating
Eq.~(\ref{eq:gpdl}) with imaginary time for the $m=1$
component only. Then the initial state of our real time
propagation is given as
$\tilde{\Psi}_m(\tilde{z})=\tilde{\psi}_g(\tilde{z})\tilde{\psi}_{\mathrm{sol},m}(\tilde{z})$,
where $\tilde{\psi}_{\mathrm{sol},m}(\tilde{z})$ is the
magnetic soliton solution or dark-bright soliton solution
in a uniform system. The two components are taken as
$m=\pm1$ and no population for $m=0$ exists. The
oscillation amplitude and period can then be determined
from the time evolution.

Since the soliton energy is given by
$\epsilon = n\hbar
    V_s\sqrt{1-U^2}$, the local density approximation (LDA)
\cite{Qu.Stringari.2016} gives the oscillation amplitude
and period of the magnetic soliton,
\begin{align}
    \frac{L}{R_z} & =\sqrt{1-(1-U_0^2)^{1/3}}, \\
    \frac{T}{T_z} & = \frac{2}{\pi}
    \sqrt{\frac{g}{\abs{\delta g}}}
    \int_0^{L/R_z} \frac{v(\beta)\dd{\beta}}
    {\sqrt{v^3(\beta)-1+U_0^2}},
\end{align}
where $R_z,~T_z$ are the Thomas-Fermi radius and
trapping period. $v(\beta) = 1-\beta^2$. $U_0$ is the
normalized soliton velocity at the center of the trap.
However, these expressions fail to predict the motion of
a magnetic soliton, as discussed in the main text.

\subsection{Experimental generation}

\begin{figure}[htbp] \centering
    \includegraphics[width=\linewidth]{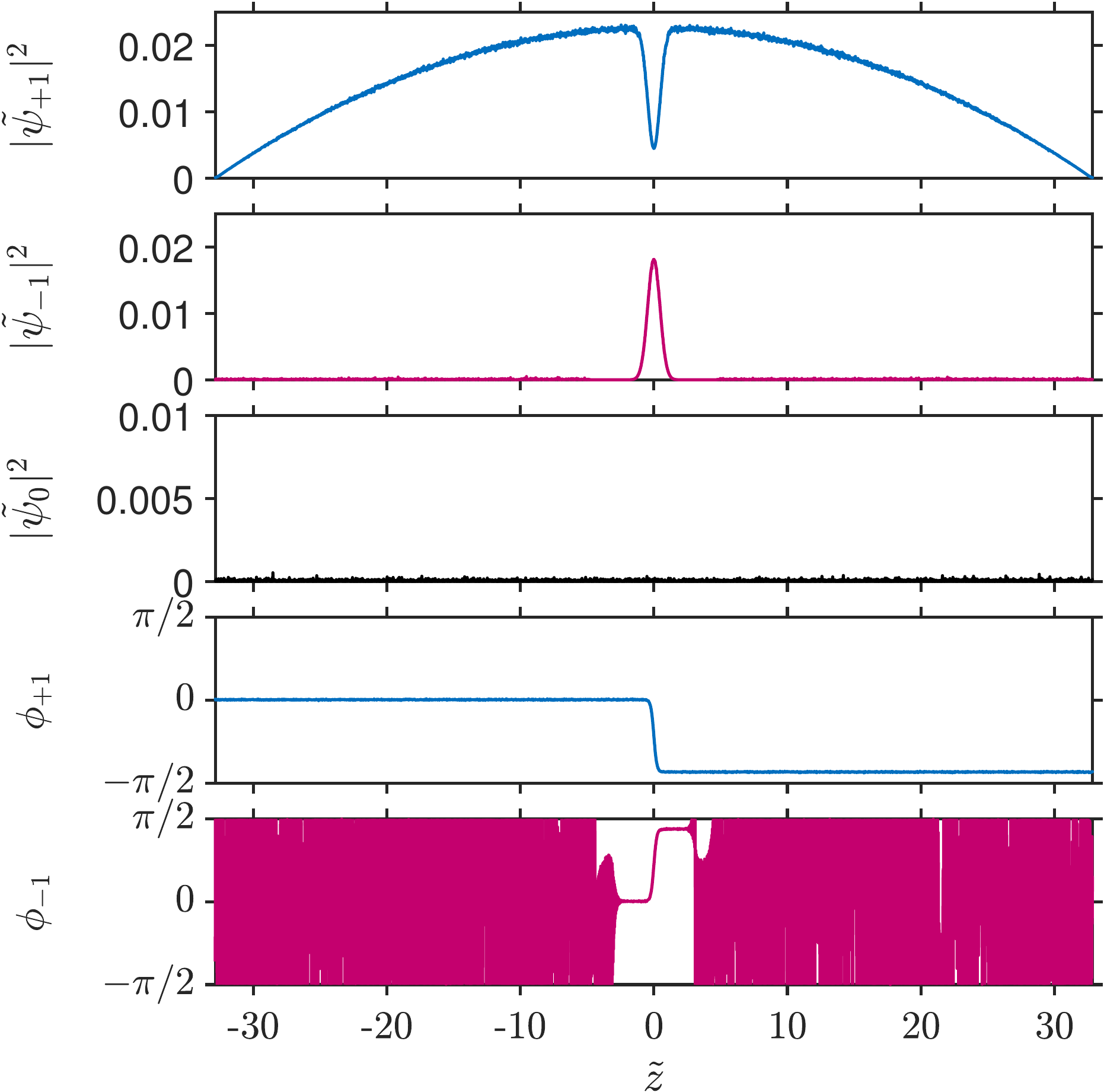}
    \caption{Initial state for generating a magnetic soliton.}
    \label{fig:sup1}
\end{figure}

As discussed in the main text, we propose to use a Raman
transition followed by a magnetic shadow (phase
imprinting) to generate a magnetic soliton in a quasi-1D
$\prescript{87}{}{\mathrm{Rb}}$ condensate. To simulate
this method, we prepare the initial condition and evolve the wave function as follows:

(1) We assume the population transfer is local and has a
Gaussian shape. We also assume that the phase imprinting
results in a $\tanh$-shaped phase step. Then the initial
state without noise is given by
\begin{equation}
    \mqty(\tilde{\varphi}_{+1}(\tilde{z}) \\ \tilde{\varphi}_{0}(\tilde{z})
    \\ \tilde{\varphi}_{-1}(\tilde{z})) = \tilde{\psi}_g(\tilde{z})\mqty(
    \sqrt{1-B e^{-\tilde{z}^2/2C^2}} e^{-i
        \phi_A(\tilde{z})} \\
    0 \\
    \sqrt{B e^{-\tilde{z}^2/2C^2}} e^{i
        \phi_A(\tilde{z})}
    ),
\end{equation} 
where $\tilde{\psi}_g(\tilde{z})$ is the
ground state wave function obtained from the imaginary
time propagation method. The phase function is given as
\begin{equation}
    \phi_A(\tilde{z}) = \frac{D}{2}(\tanh{\frac{\tilde{z}}{E}}+1).
\end{equation}
In our simulation we use the following dimensionless
parameters,
\begin{equation}
    B = 0.8,~C = 0.463,~D = 1.37,~E = 0.216.
\end{equation}
For comparison, the dimensionless spin healing length
in our simulation is $\tilde{\xi}_s = 0.316$ evaluated 
at the center of the condensate. 

(2) Then we include noise to the initial condition as
\begin{align}
    \tilde{\psi}_{+1}(\tilde{z}) & = 
    \tilde{\varphi}_{+1}(\tilde{z}) 
    \{ 1+\eta_1(\tilde{z})+i\eta_2(\tilde{z})\}, \\ 
    \tilde{\psi}_{0}(\tilde{z}) & = 
    \tilde{\varphi}_{0}(\tilde{z}) +
    \eta_3(\tilde{z})+i\eta_4(\tilde{z}),\\
    \tilde{\psi}_{-1}(\tilde{z}) &= 
    \tilde{\varphi}_{-1}(\tilde{z})
    \{ 1+\eta_5(\tilde{z})+i\eta_6(\tilde{z})\} \nonumber \\
    &\hspace{1.5cm}+\alpha(\tilde{z})\{ \eta_7(\tilde{z})+i\eta_8(\tilde{z}) \},
\end{align}
where $\alpha(\tilde{z}) = 0$ for $-5<\tilde{z}<5$ and 
$\alpha(\tilde{z})=1$ otherwise. $\eta_j(\tilde{z})$ is
Gaussian noise sampled with the standard deviation 0.005.
The initial density distributions and phase profiles are
shown in Fig.~\ref{fig:sup1}.

\begin{figure}[htbp] \centering
    \includegraphics[width=\linewidth]{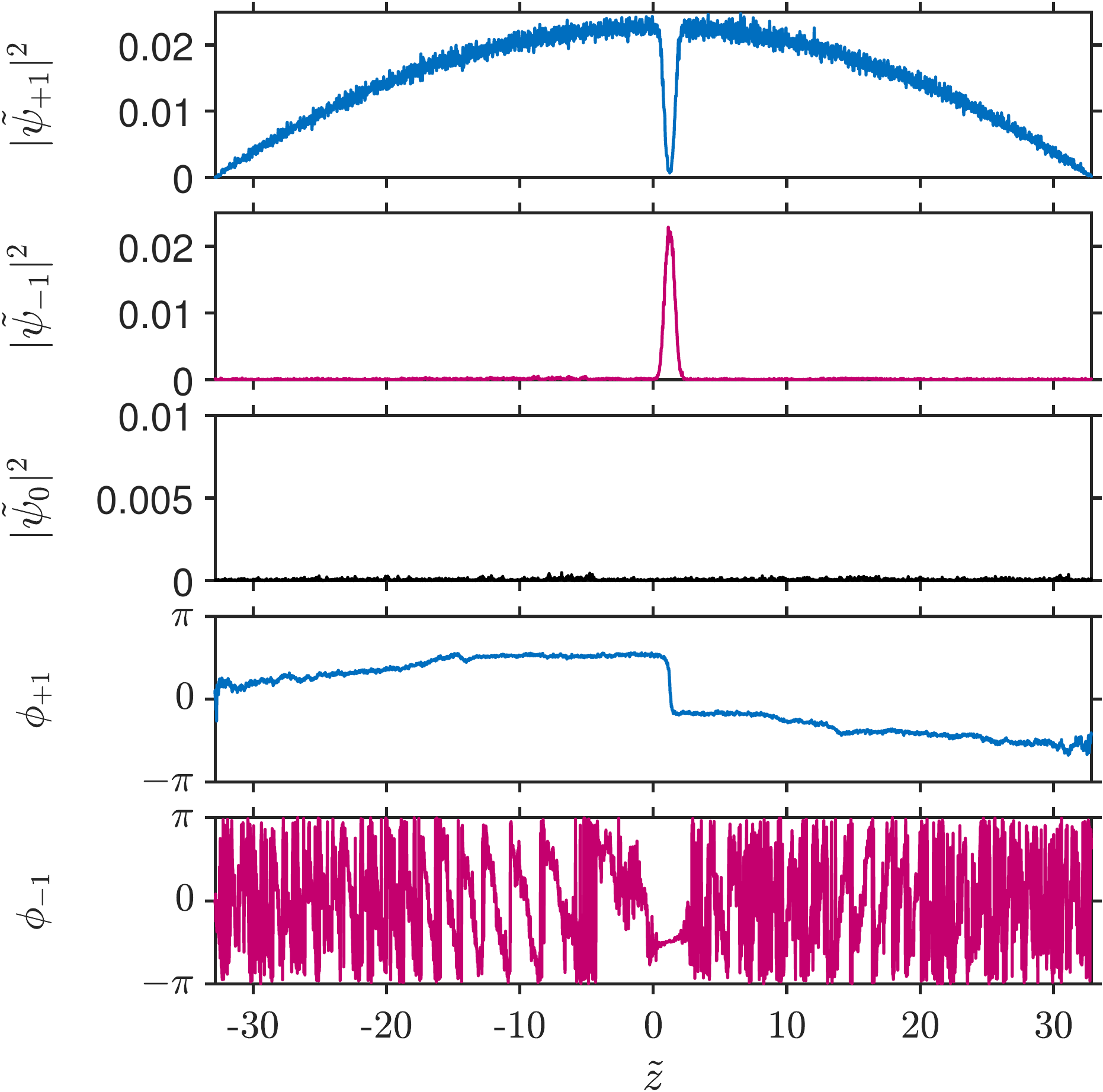}
    \caption{Density and phase profiles at $\tilde{t} = 3.77$.}
    \label{fig:sup2}
\end{figure}

(3) We then numerically solve Eq.~(\ref{eq:gpdl}) with 
time step $\Delta \tilde{t} = 1.885\times 10^{-4}$. A
quadratic shift $\tilde{q} = -10$ is added to stabilize
the condensate. The resultant magnetic soliton resembles
the ideal case of no noise as shown in
Fig.~\ref{fig:sup2}.

\end{document}